# Atomic Layer Deposition Nucleation Dependence on Diamond Surface Termination

*Jessica C. Jones[1], Nazar Delegan[1,2], F. Joseph Heremans[1,2,3], Alex B. F. Martinson,[1,*]*

[1]Materials Science Division, Argonne National Laboratory, Lemont, IL, 60439, USA
[2]Center for Molecular Engineering, Argonne National Laboratory, Lemont, IL 60439, USA
[3]Pritzker School of Molecular Engineering, University of Chicago, Chicago IL 60637, USA

Abstract

Surface termination and interfacial interactions are critical for advanced solid-state quantum applications. In this paper, we demonstrate that atomic layer deposition (ALD) can both provide valuable insight on the chemical environment of the surface, having sufficient sensitivity to distinguish between the common diamond (001) surface termination types and passivate these interfaces as desired. We selected diamond substrates exhibiting both smooth and anomalously rough surfaces to probe the effect of morphology on ALD nucleation. We use high resolution *in situ* spectroscopic ellipsometry to monitor the surface reaction with sub-angstrom resolution, to evaluate the nucleation of an ALD $Al_2O_3$ process as a function of different ex and in situ treatments to the diamond surface. In situ water dosing and high vacuum annealing provided the most favorable environment for nucleation of dimethylaluminum isopropoxide and water ALD. Hydrogen termination passivated both smooth and rough surfaces while triacid cleaning passivated the smooth surface only, with striking effectiveness.

Highlights

- Modification of diamond surface accomplished in situ prior to thin film deposition.
- Diamond termination significantly impacted nucleation behavior of atomic layer deposition.
- ALD is used as an analytical tool to evaluate diamond surface termination and morphology.

[*] Corresponding author. Tel: 630 252-7520. E-mail: martinson@anl.gov (Alex Martinson)

**Graphical Abstract**

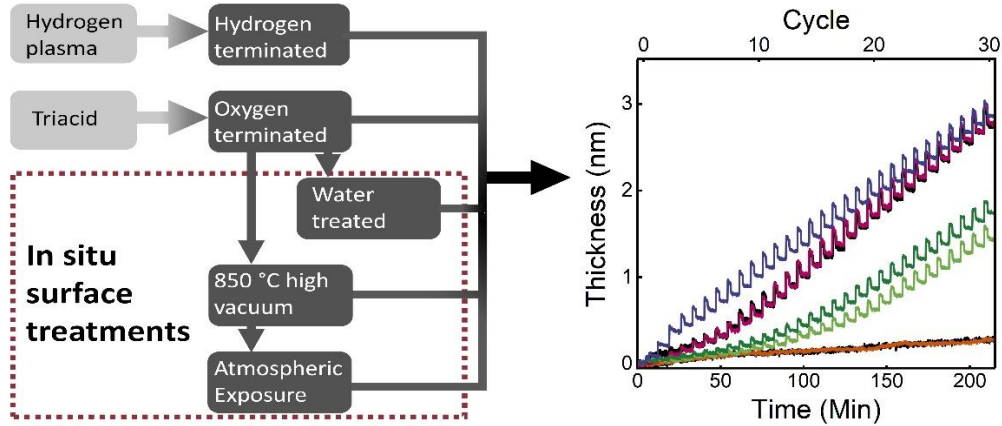

1. Introduction

Understanding and engineering diamond surfaces and interfaces is critical for both electronic [1,2] and quantum applications [3–6]. Quantum relevant spin defects in diamond, such as the negatively charged nitrogen-vacancy (NV) center, are sensitive to, amongst other things, their host's band structure, charge environment, and magnetic environment [6–8]. In this context, interfaces act as intrinsic discontinuities, drastically affecting the nearby defects' charge state, optical stability, and coherence times [3,5,7]. These challenges are further exacerbated for applications related to quantum sensing and device integration [9–11] where there is a strong advantage to localizing the spin defects near (<50 nm) or at the host's surface/interface. In response, much effort has been placed in deterministically and homogeneously terminating diamond's surface [3,5,6,12]. A uniformly hydrogenated diamond surface is desirable for some applications such as the formation of two-dimensional hole-gases [13,14]. In the context of NV-based quantum sensing, hydrogen terminated surfaces result in negative electron affinity, undesirable pinning of the Fermi level, and a deep depletion region, which can result in the quantum relevant negative NV charge state ionizing into a dark NV0 state [5,7].

When left untreated and exposed to atmosphere, diamond (001) surfaces will adopt mixed termination states, dictated by environmental composition, resulting in heterogeneous and variable surface charge environments and dangling bonds that lead to unpredictable spin, charge, and coherence environments [3,5]. As such, several surface species (e.g. hydroxyl, carbonyl, nitrogen, ether, etc.[4,5]) have been investigated in an effort to create low decoherence, charge stabilizing interfaces for solid state spin qubit integration. Well-ordered, primarily ether-terminated diamond surfaces are desirable for quantum applications, with simple triacid cleaning produces a less homogenous surface than annealing in oxygen [3]. While annealing in oxygen produces well-ordered surfaces it can be time consuming and require surface characterization studies during the process. As such, practical and rapid approaches to both elucidate and terminate the desired diamond surface state remains elusive. Atomic layer deposition (ALD) utilizes volatile and reactive chemical precursors to deterministically direct surface chemistry and precisely deposit myriad materials from across the periodic table. This makes ALD a compelling approach to engineer interfaces due in part to moderate-temperature and scalable processing, the potential for chemical specificity, the wide range of accessible chemistries and exquisite control of thickness [16,17]. Most reports of ALD on diamond to date focus on an H-terminated surface for semiconducting applications [1,2]. Only recently has ALD been investigated to passivate diamond surfaces for quantum sensing applications, for example for deposition of a thin passivation layer that resulted in minor changes to coherence times for sensing of biomolecules [18,19].

In contrast to many deposition techniques, ALD relies solely on chemical reactions with the substrate surface to initiate film growth. Therefore, substrate surface chemistry, morphology, reconstruction, and termination are critical to understanding and controlling ALD reactivity, and by extension, nucleation [20–22]. Detailed nucleation studies on diamond surfaces are lacking, which may explain the varying results observed for ALD of dielectric films on diamond [23,24]. A study comparing ALD alumina ($Al_2O_3$) on oxygen plasma treated (oxygen terminated) and as grown (hydrogen terminated) polycrystalline diamond surfaces observed enhanced adhesion of the film on the oxygen terminated surfaces. While delamination was observed on the hydrogen terminated surface, favorable reactions between the ALD precursors and "oxygen terminated" surface chemistry (-OH and C-O-C) was hypothesized to improve adhesion [24]. These observations highlight the need for further investigation of ALD of diamond surfaces.

In this work, we aim to address the various challenges presented above by uniquely combining deterministic synthesis of near surface NV centers in diamond, analytical ALD, and in-situ ellipsometry to demonstrate the applicability of ALD to both study and control the surface of (001) diamond. Specifically, we aim to investigate the sensitivity of ALD nucleation as a function of quantum-relevant diamond surface termination schemes for the identification of (001) surfaces. In addition, we investigate how ALD can be used to passivate diamond surfaces, paving the way for long-term stability and spin coherence studies in the future.

2. Methods

2.1 Sample preparation

All substrates studied are optical grade diamond (Element Six, 3x3x0.3mm). These substrates were He implanted and thermally processed, creating a ~370 nm deep graphitization layer [25]. The substrates were then overgrown and processed to create a delta doped NV layer ~25-50nm from the surface [26]. These NV centers were synthesized for future coherence measurements as a function of surface termination. As such, these same samples can be used for both classical and quantum characterization. Two samples were evaluated in this work, one showing rougher surface features, presenting an average RMS of ~8.3 nm and showing step flow grow with nanometer scale terrace pinning, with a second sample showing classical smooth step flow features with an average RMS of ~0.4 nm (see figure 1). The samples were used to de-couple the effect of surface roughness on the surface nucleation studies presented herein.

The diamond surface roughness and morphology were characterized via using a Bruker Dimension FastScan AFM, with both large (>10 um) and small (~500 nm) scan areas. Hydrofluoric acid (10% buffered oxide etchant) was used before the initial ALD growth to remove any graphitic carbon, and between ALD growths to clean the surface (30 min dip with double DI rinse) and prepare the diamond substrates for subsequent surface re-termination processing. The surface morphology and roughness as measured by AFM did not significantly change with HF etching of the diamond surfaces between experimental conditions.

2.2 Ellipsometric measurements and modelling

The subsurface graphitization layer caused by the helium implantation produces strong optical contrast with the surrounding diamond that allows measurement with spectroscopic ellipsometry (SE). We introduce SE as a novel route for the rapid, non-destructive measurement of graphitization layer depth in diamond with high precision. In situ and ex situ spectroscopic ellipsometry were acquired from 245 to 1000 nm (J. A. Woollam Co, M-2000X) and modeled with CompleteEASE software. The ellipsometric models consisted of diamond substrate, a ~13 nm thick graphite layer, and a diamond (GenOsc) material overlayer. The thickness of the ALD aluminum oxide deposition was modeled with a Cauchy model with parameters locked to that of a thick ALD $Al_2O_3$ film previously grown on a Si wafer using the same process conditions.

The adjustable parameters of both diamond layers, as well as the thickness of the graphitic and the diamond overlayers were initially fit to match the sample, then locked in further studies. This model proved accurate for the measurement of the graphitic layer and diamond overgrowth layer. These values were compared to SIMS characterization of CVD diamond growth rate, confirming the overgrown total diamond thickness (~500 nm) above the ~13nm thick graphitic layer [25]. A set of previously reported diamond surface termination strategies were applied via a combination of wet and dry processes. Nominally hydrogen terminated surfaces were achieved via $H_2$ plasma exposure of the samples inside a SEKI CVD reactor at 25 Torr, 400 sccm $H_2$ flow, 900W micro-wave power for 20 min [25]. Nominally oxygen terminated surfaces were achieved via triacid cleaning, i.e. a slow-ramping boil of 1:1:1 nitric, sulfuric, and perchloric acids heated to <250 °C (below the sulfuric boiling point) [27]. Further modification of surface chemistry was achieved via in-situ processing as illustrated in Figure 1.

2.3 ALD

ALD relies on cycled self-limiting surface reactions that result in sub-nanometer layer-by-layer growth after nucleation. $Al_2O_3$ was grown by alternating cycles of dimethyaluminum isopropoxide (DMAI, >99.999% purity [Wonik]) and ultrapure water (18MΩ, [Millipore Inc.]). DMAI was loaded into a 50 mL stainless steel cylinder (Swagelok) under nitrogen in a glovebox and ultrapure water was loaded into a $2^{nd}$ stainless steel cylinder. Both precursors were delivered under their own vapor pressure, with the DMAI cylinder heated to 65 °C and no heating of the water cylinder. ALD processing was performed under continuous 500 sccm ultrapure Ar flow through two 3-way ALD valves (Swagelok). Ultrapure Ar gas used as the carrier gas was further purified by an inert gas purifier to <1 ppt $O_2$ and $H_2O$ (Entegris gatekeeper) just prior to entering the growth chamber. A custom-built high vacuum/ALD tool was used for both the ALD experiments and the high vacuum annealing. DMAI and water vapor ALD process conditions were optimized for growth on itself (see SI), and a dosing schedule was selected that applies approximately twice the saturating dose of both precursors (0.25 – 120 – 0.3 – 300 s; DMAI-purge-$H_2O$-purge). Saturating steady-state growth of $Al_2O_3$ was measured to be 0.1 nm/cycle at the 150 °C substrate temperature that was utilized for all ALD growths.

3. Data and Results

The stability of diamond surfaces is paramount to the reproducibility of nearby NV-centers for quantum sensing applications. Oxygen plasma, UV ozone, and triacid cleaning have previously been applied to as-grown diamond to "oxygen terminate" the substrate [5]. Diamond treated with triacid cleaning present a less ordered and more inhomogeneous mix of states compared to oxygen annealed surfaces when studied with NEXAFS [3]. Possible inhomogeneities in the oxygen bonding environment produced by triacid cleaning have been proposed to contribute to a higher density of unoccupied states below the conduction band which can act as electron traps contributing to magnetic noise and lowering coherence times, however there is some disagreement on the ideal chemical environment of the surface should be to maximize coherence [3,6]. The long-term optical properties of a triacid cleaned surface was observed to change when monitored by ellipsometry for > 6 weeks (see SI). The substrate, graphitic layer, and diamond overlayer optical properties and thicknesses were locked and a transparent overlayer was used to model the near-surface. The apparent thickness of this layer trended upward (~0.05 Å/week) and fluctuated with ambient conditions (i.e. humidity), further motivating the understanding of the surface evolution in time and the use of protective overlayer formation through ALD.

ALD films have been studied as passivation barriers previously, primarily using alternating cycles of trimethylaluminum (TMA) and water on H-terminated surfaces where $Al_2O_3$ stabilizes this H-diamond hole channel by maintaining the hole sheet concentration for air and NO$_2$ adsorbed surfaces [2,28–30].  Detailed studies on the nucleation of ALD films on diamond remain sparse [24,31].  In a report of ultra-thin (0.5-3 nm) ALD Al$_2$O$_3$ on triacid cleaned diamond, static water contact angle was used to determine when the ALD film had coalesced to provide a suitable surface termination for subsequent formation of a self-assembled monolayer (SAM).  Characterization of nucleation was confirmed solely via the thickness of the resulting film from AFM on a section of film removed [19]. The strong Lewis acidity and ready reactivity of TMA allows for nucleation on diverse surfaces via numerous mechanisms to allow deposition even on nominally carbon-presenting substrates including self-assembled monolayers and graphene [32,33].

While the ALD Al$_2$O$_3$ process alternating TMA with water is robust, the extensive reactivity often results in non-specific and heterogeneous binding with a notable lack of surface selectivity [32].  The process may be useful for many applications in which high nucleation density is desired (e.g. pinhole-free barrier layers), however TMA can chemically alter some substrates [34] and provides little chemical insight into the diamond surface present under ALD process conditions [35].  Therefore, we investigate the ALD nucleation and growth of Al$_2$O$_3$ via an alternative precursor, namely dimethylaluminum isopropoxide (DMAI), after considering recent reports of increased selectivity, especially for proton-mediated ligand exchange reactions with hydroxyl-terminated surfaces [36,37]. There is now strong evidence that DMAI reacts selectively with hydroxyl ligands on single crystal TiO$_2$ surfaces resulting in a large nucleation delay when no surface hydroxyls are available with which to initiate deposition [38,39].  This selectivity, in turn, presents an attractive method to probe hydroxyl surface termination on otherwise unreactive surfaces including diamond.  A schematic of the experimental design is presented in Figure 1.

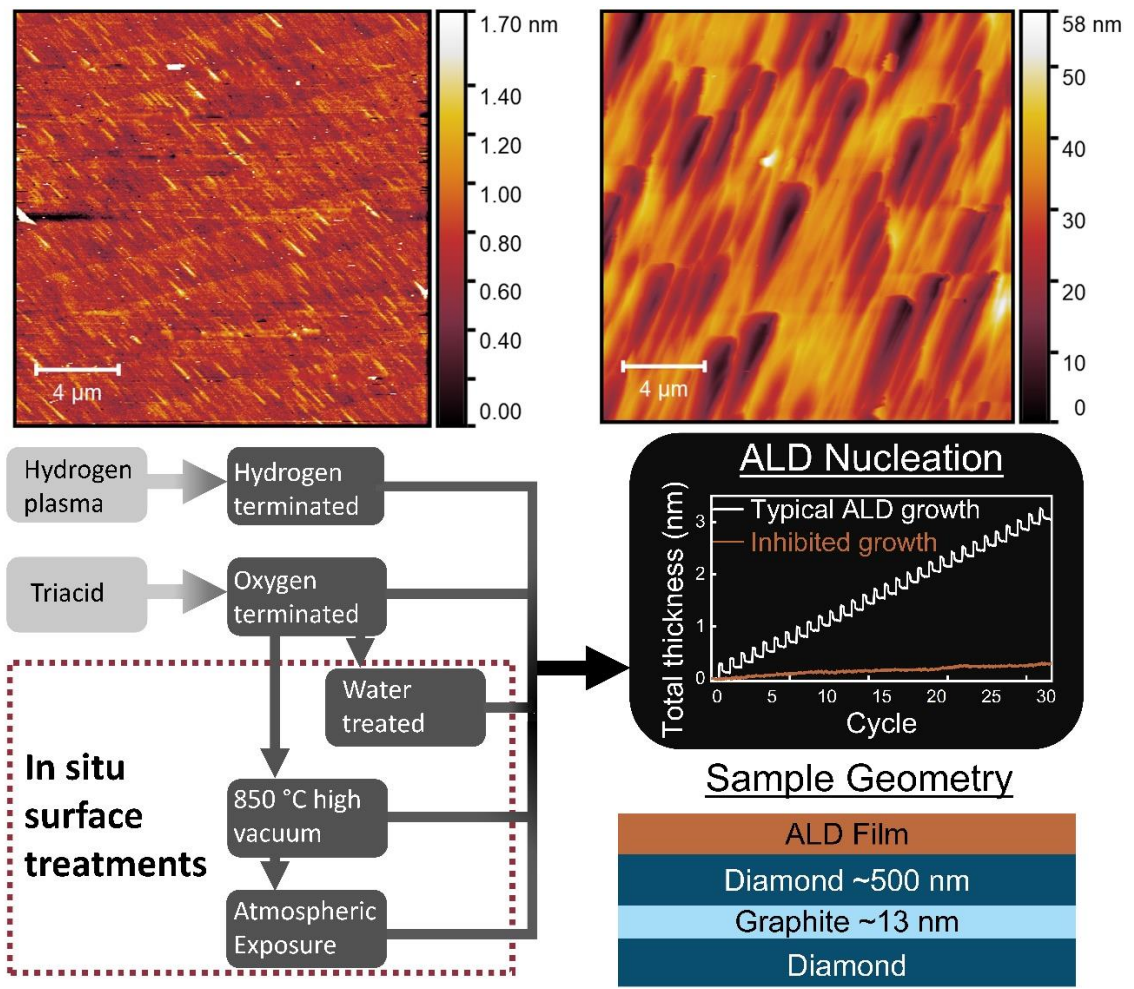

**Figure. 1** (top) AFM from both the smooth (top-left) and rough (top-right) diamond surfaces. The smooth surface presents clear step flow features with an RMS ~ 0.4 nm, whereas the rougher sample shows step flow features with terrace pinning features and an RMS of ~ 8.3 nm. (bottom) Experimental workflow of surface treatments, diamond sample geometry, and examples of typical, and inhibited ALD nucleation. Sample geometry is not to scale.

After ALD nucleation is tested on each termination of the surface, the oxide layer is etched off with 10% HF content, buffered oxide etchant (BOE) to reset the surface prior to the next termination. In this way much of the sample-to-sample variation is reduced as both the rough (8.3 nm RMS) and smooth (<0.4 nm RMS) crystals are subjected to each treatment and compared. We begin by investigating nucleation on H-terminated diamond surfaces as produced by CVD diamond growth or via dry H-termination. We observe no clear nucleation of ALD $Al_2O_3$, as would be indicated by a regular stair-step thickness increase, on H-plasma treated diamond surface for rough or smooth samples, Figure 2. This result is consistent with hydrogen termination that is agnostic to the nature of the exposed crystallographic facets for the rough sample. The lack of nucleation corroborates the absence of hydroxyl groups or other surface species reactive to DMAI.

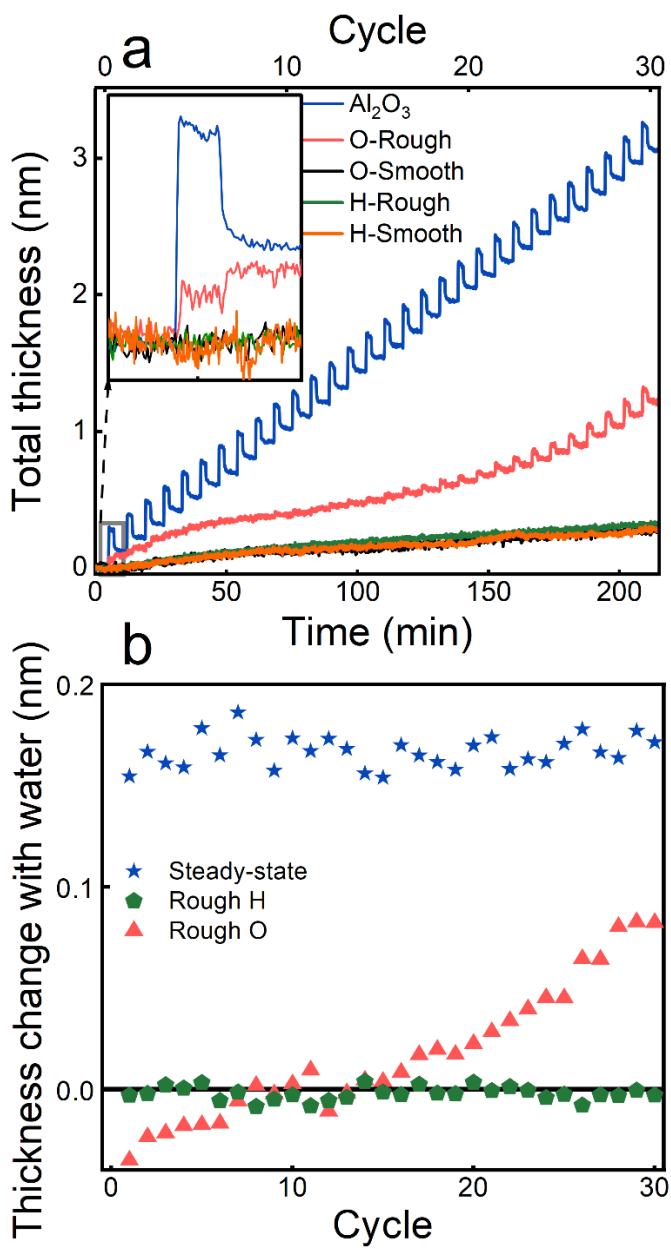

**Figure. 2** (a) Total optical thickness of $Al_2O_3$ measured during $DMAI/H_2O$ ALD on $Al_2O_3$ (steady-state growth) and on rough and smooth diamond substrates that have been freshly triacid, or H-plasma treated. (b) Optical thickness change upon water exposure for each ALD cycle of rough H plasma treated and triacid cleaned diamond.

The in situ ellipsometric measurement of an oxide overlayer on diamond with underlying graphitized interlayer exhibits thickness precision better than ~0.02 nm, more than an order of magnitude below monolayer thin film coverage (~0.3 nm). This allows for detection of the earliest stages of $Al_2O_3$ nucleation to better than 7% of a complete monolayer. The absence of ALD $Al_2O_3$ deposition coincident with DMAI or $H_2O$ exposure on smooth or rough H-terminated diamond is a remarkable result as very few air-exposed surfaces show such strong ALD inhibition, suggesting the H-terminated surface is highly inert to these ALD growth conditions.

Strikingly, nucleation on the smooth triacid treated surface is equally inhibited. While there is strong evidence for oxygen containing surface chemistry of triacid treated diamond including but perhaps not limited to C=O, C-O-C, or C-O-H [3] the precise surface chemistry remains unknown. One commonly proposed mechanism for metal oxide ALD processes is the reaction of a hydroxylated surface site with the metalorganic precursor, here DMAI. A simple proton exchange affords binding of the metal to surface oxygen with the release of a protonated ligand, here methane or isopropanol. The strong inhibition of ALD $Al_2O_3$ that we observe suggest that surface hydroxyls (-OH) are not present or that DMAI reacts slowly or is inert to -OH terminated diamond at 150 °C. In contrast to the H-terminated diamond surfaces, the rough triacid treated diamond surface results in delayed, yet clear nucleation of $Al_2O_3$ with DMAI and $H_2O$. This may be due to more diverse surface termination induced by triacid treatment of more varied facets and edges exposed by the rough diamond substrate. It has been predicted and shown that distinct facets of diamond alter the extent and type of oxygen termination present [5]. This facet dependence is also observed in oxide substrates including $TiO_2$ and $In_2O_3$, which have been computationally investigated in detail to predict step edges surface chemistry that is distinct from the terraces [20,40]. Therefore, we hypothesize that a rough diamond surface is prone to molecular or dissociative absorption of water after triacid treatment. This hypothesis is further consistent with the ellipsometric fluctuations observed under variable humidity, (see SI). The exact transition between the smooth and rough surface regimes remains unclear and will be the subject of subsequent studies.

An analysis of the optical thickness change within each ALD cycle provides an even more sensitive measurement of nucleation, Figure 2b. For example, post-nucleation the $Al_2O_3$ ALD growth process exhibits a regular optical thickness increase upon exposure DMAI and partial decrease upon $H_2O$ exposure (see inset of figure 2a and SI). Monitoring this thickness oscillation enables even greater precision as each synchronized step is less susceptible to the effects of background measurement drift (e.g. from the slow evolution of sample temperature and/or position). The change in optical thickness produced during steady-state growth of $Al_2O_3$ during water dosing ($\Delta H_2O$) reveals steady removal of DMAI ligands from the surface equivalent to 0.17 nm/cycle (see Figure 2b). In contrast, we observe no clear steps in synchrony with ALD precursors dosing for rough H-terminated diamond. This analysis corroborates the near absence of ALD growth/nucleation over 30 cycles of ALD processing despite a very slow increase in the apparent thickness with time (~0.1 nm/hr).

The same $\Delta H_2O$ analysis of the rough triacid treated diamond substrate reveals a negative $\Delta H_2O$ for the first 7 cycles that is not consistent with surface-bound-DMAI ligand removal but instead with direct reaction of water with the surface. After 7 ALD cycles $\Delta H_2O$ becomes positive and continues to increase with cycle number, signifying that the expected surface-bound-ligand removal may outweigh any further $H_2O$ reaction with the rough diamond surface. This analysis is consistent with a measurable increase in $\Delta$DMAI shortly after 7 cycles, see SI. The larger growth profile is consistent with island growth, as modeled by Nilsen et al, where the precursor adsorbs to the surface and the island grows vertically and laterally [41]. However, this profile could also be consistent with the model proposed by Parson et al where progressively more nucleation sites are generated during the ALD process, possibly due to the exposed diamond surface's susceptibility to further reaction with water at 150 °C [42]. In short summary, we find that atomically smooth single crystal diamond substrates are remarkably resistant to ALD nucleation with DMAI and $H_2O$ at 150 °C.

For completeness, we evaluated the effect of in situ surface treatments on a nominally 'clean' (i.e. triacid treated) diamond surface. In situ treatments included water exposure at elevated temperature [43], annealing in high vacuum at 850 °C, and exposure of vacuum annealed samples to atmospheric contamination. Both smooth and rough diamond surfaces exhibited rapid nucleation, under the same ALD processing conditions as those previous applied, after exposure to water at 250 °C, Figure 3. While nucleation was rapid on both surfaces, it is notable that this was the only example in this study where the smooth surface exhibited slightly more rapid nucleation than the rougher substrate. One possible explanation is that the smooth substrate was more uniformly -OH terminated upon elevated temperature water exposure while the rougher surface presented more inhomogeneous surface chemistry that was less reactive with these ALD precursors. This dramatic change in nucleation of $Al_2O_3$ ALD films on the diamond surface after mild heating and water exposure suggests a significant alteration of the diamond surface chemistry that provides a simple route to more reliable ultra-thin film growth and more uniform composition/connectivity between diamond and the oxide overlayer that merits further study.

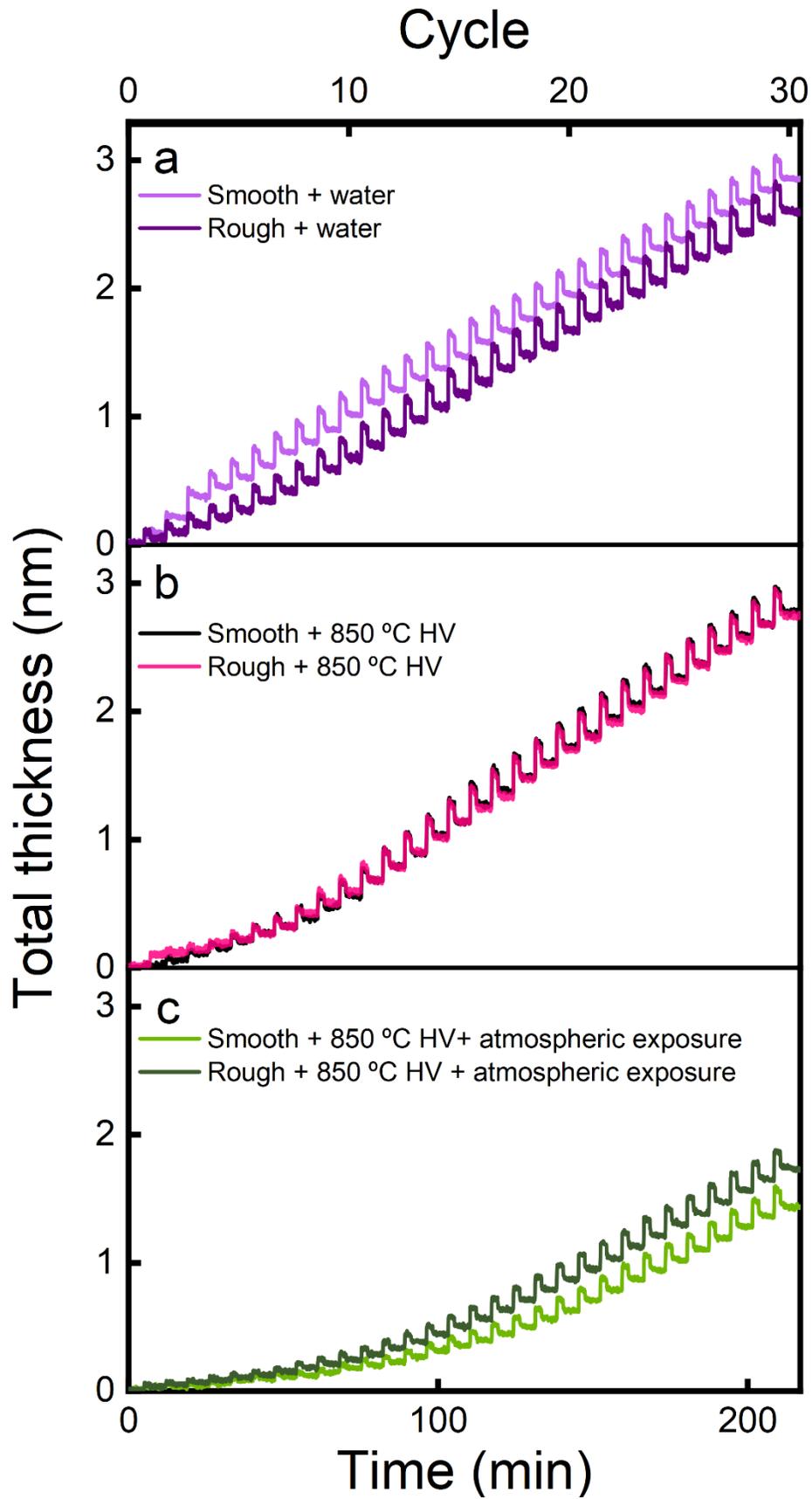

**Figure 3.** Thickness of $Al_2O_3$ deposited at 150 °C on triacid treated diamond upon ALD processing after (a) 30 pulses of water at 250 °C, (b) annealing to 850 °C in high vacuum, (c) annealing to 850 °C and subsequent exposure to air.

Annealing in vacuum has been reported to remove adsorbed water at low temperatures and remove the ketone-like bound oxygen from the surface at higher temperatures [44,45]. Both smooth and rough triacid treated diamond substrates were introduced to the ALD growth chamber and annealed to 850 °C in high vacuum, base pressure $1\cdot10^{-8}$ Torr. This treatment has been referred to as a "reset" of the diamond surface with the intention of creating a carbon terminated surface by removing the oxygen from the surface [3]. However, after annealing triacid treated diamond samples, both rough and smooth, we observe relatively rapid nucleation upon ALD processing, with both samples reaching the steady state growth rate (0.1 nm/cycle) after 10 cycles, Figure 3. The change in nucleation behavior upon in situ treatment is unlikely to be correlated with a change in surface morphology as AFM images acquired after annealing and deposition revealed no major change in surface roughness (see SI). While the precise surface chemistry of the surface remains unknown, we note that ALD nucleation is significantly delayed relative to a smooth surface that has been exposed to water at 250 °C.

To investigate the stability of high temperature vacuum annealed diamond surface the same triacid treated-high vacuum annealed samples were exposed to air for several hours (smooth surface for 4 hours, rough surface for 48 hours) before being reintroduced to the ALD chamber for nucleation. These annealed and atmosphere exposed samples showed less rapid nucleation than those that did not experience a vacuum break. This indicates that some portion of the sites created by annealing that enable prompt nucleation were altered via interaction with the atmosphere, which indicates that the surface still exhibits some inhomogeneity. Therefore, we note that the "shelf-stability" of this in-situ treatment is lacking, further highlighting the need for a quantum-compatible, shelf-stable approach to diamond surface termination.

An overview of ALD nucleation dependence on diamond surface termination may be gleaned from quantification of a "nucleation delay" as calculated by the x-intercept of the linear regression of the thickness versus cycle data once the growth has reached steady state (0.1 nm/cycle). A compilation of this analysis, Figure 4, reveals a range of nucleation delays that depend primarily on surface treatment and secondarily on substrate roughness.

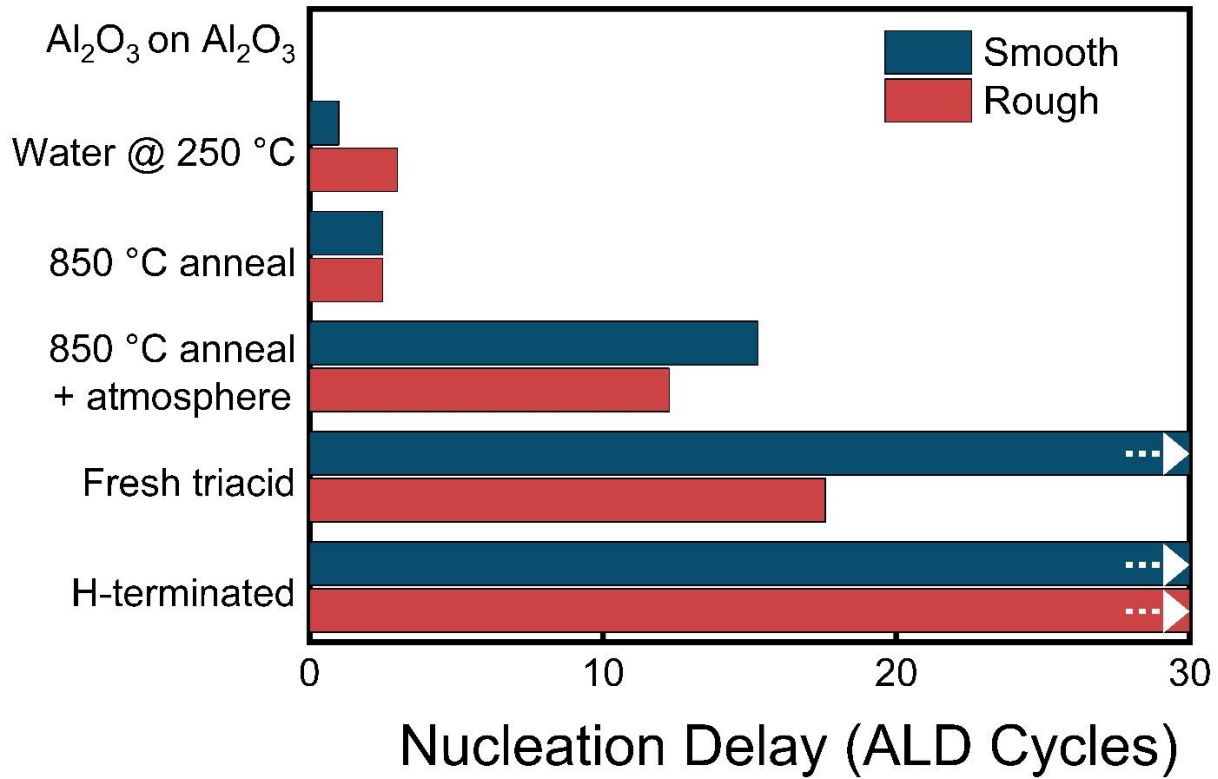

**Figure 4.** Nucleation delay (measured in ALD cycles) as a function of diamond surface preparation. Arrows indicate the true nucleation delay is greater than 30 cycles for samples the samples that did not nucleate during the experiment.

The hydrogen plasma treated and triacid cleaned were the most resistant to nucleation under these ALD processing condition, whereas the water treated, and high vacuum annealed surfaces were the least resistant to ALD. This brief summary of the effect of surface treatments on nucleation of ALD grown $Al_2O_3$ using DMAI and water demonstrates the potential to use ALD as an analytical technique to probe surface reactivity as well as guides the choice of diamond surface treatments in preparation for application in quantum technologies.

4. Conclusions

We demonstrate the applicability of ALD as an analytical tool to study quantum technology relevant diamond surface terminations and states. We demonstrate clear distinguishability between the commonly reported diamond surface termination schemes [5] via their nucleation susceptibility during $Al_2O_3$ ALD. Furthermore, we demonstrate that ALD nucleation can be used to investigate the (in)homogeneity of diamond surfaces and that surface passivation and re-termination can be accomplished in situ, paving the way for quantum-relevant control and stabilization of diamond surfaces. Subsequent studies will focus on long-term NV coherence characterization of the surface spin bath noise environment as a function of termination and ALD-based passivation schemes.

**Acknowledgment**

This work was supported by the U.S. Department of Energy, Office of Science, Basic Energy Sciences, Materials Science and Engineering Division (J.C.J., N.D., F.J.H, A.B.F.M.). Work performed at the Center for Nanoscale Materials a U.S. Department of Energy Office of Science User Facilities, was supported by the U.S. DOE, Office of Basic Energy Sciences, under Contract No. DE-AC02-06CH11357. All authors confirm that there are no conflicts of interest with this article.

**Supporting Information**

The Supporting Information is available free of charge on the internet.

# Supporting Information

**Experimental**

Custom Built ALD Tool

High vacuum annealing and ALD processing were performed in a custom-designed, 12" diameter spherical vacuum chamber. Samples were secured to Ta flags with Ta foil straps and seated on the sample stage adjacent to a resistively heated pyrolytic boronitride heater (Ferrovac). High vacuum annealing was performed with chamber walls heated, and the pyrolytic boronitride heater temperature adjusted to temperature. Temperatures were recorded based on a k-type thermocouple mounted near the sample on the heating stage. High vacuum annealing was performed under turbomolecular pumping with chamber base pressure approximately $2 \times 10^{-7}$ Torr. ALD processing was performed under rough mechanical pumping with sample stage and chamber walls heated to 150 °C. Ultrahigh purity argon purge gas was further purified to <1 ppt $O_2$ and $H_2O$ with an inert gas filter (Entegris, Gatekeeper) and introduced through two three-way ALD valves at a total flow of 500 sccm, resulting in a base pressure of 0.9 Torr. Sample exchanges are performed through gated load lock chamber, after which the sample is allowed to equilibrate at the growth temperature, Ar flow, and vacuum conditions for 2 hours prior to initiation of ALD processing. In situ spectroscopic ellipsometry is performed using ports at 70 ° from surface normal, with the focal point on the sample stage.

The nucleation density of $Al_2O_3$ on diamond surfaces was deduced from fitting the ellipsometric thickness after each full ALD cycle to a random island nucleation model developed by Nilson et al.[1] This model assumes that nucleation occurs only in the first cycle and that the nuclei expand through ALD growth as a hemisphere at a growth rate equal to the thin film growth rate, aka the intrinsic ALD growth rate. Once the islands (with radius r) grow large enough to exceed the radius of convergence $R_{cov}$ they coalesce into a film slightly rougher than the substrate.

1.1 $\mu =$

$$\begin{cases} \left(\frac{2}{3}\right) N_d \pi r^3 & r \leq R_{cov} \\ N_d \left(\pi R_{cov}^2 \sqrt{r^2 - R_{cov}^2} + \left(\frac{\pi}{6}\right)\left[3R_{cov}^2 + (r - \sqrt{r^2 - R_{cov}^2})^2\right](r - \sqrt{r^2 - R_{cov}^2})\right) & r > R_{cov} \end{cases}$$

1.2 $R_{cov} = \sqrt{1/\pi N_d}$

1.3 $r = gx$

**Equation 1:** 1.1 Simplified Island nucleation model adapted from Nilsen et al.[1] 1.2 is the radius of coverage as defined for equation 1.1 with unit cells simplified to a disk from a hexagon. 1.3 equation for the radius of each island assuming growth in all direction equal to steady state growth here (0.1 nm) multiplied by the cycle number (x).

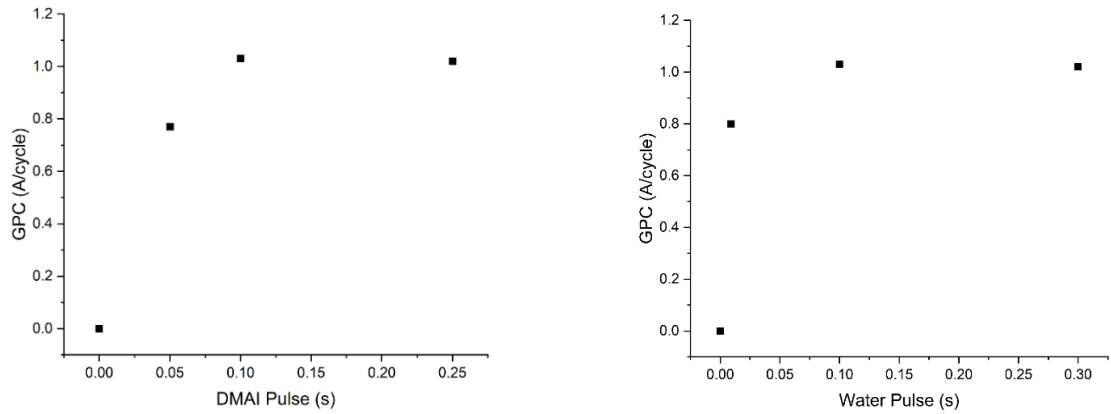

**Figure. S1.** Saturation curves of $Al_2O_3$ ALD on itself. Thickness of $Al_2O_3$ deposited by DMAI/$H_2O$ ALD process at 150 °C as deduced from in situ spectroscopic ellipsometry. During water saturation studies the DMAI pulse time = 0.25 s, while during the DMAI saturation study the water pulse time = 0.3 s.

The stability of freshly triacid-treated CVD grown diamond samples was monitored by spectroscopic ellipsometry in ambient lab environment for > 6 weeks. A model comprising bulk diamond/graphite (13.5 nm)/diamond (512.8 nm) fit well to ellipsometric data collected immediately after triacid cleaning. Improved fitting was achieved through addition of a nominally transparent ultrathin film of moderate index modelled as a topmost Cauchy layer with parameterized constants (A=0.1 B=0.19767 C=0) and fitted thickness. The parameters of diamond substrate, interlayer, diamond overlayer and transparent Cauchy film were locked based on the initial measurement such that only the thickness of the Cauchy was allowed to change over the course of the experiment. The apparent thickness of this thin film increases steadily over the course of the experiment, **Figure S2**.

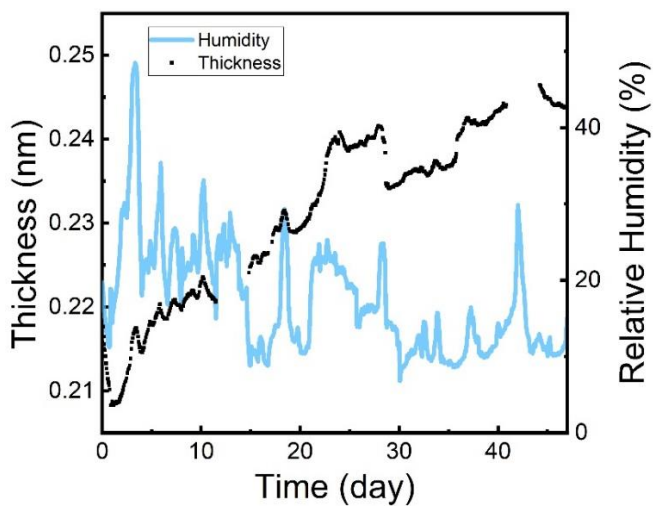

**Figure. S2.** Apparent thickness of overlayer on triacid cleaned diamond, (left axis) and raw relative humidity measured inside the building (right axis).

Humidity fluctuations within the building are clearly correlated to the apparent thickness of the transparent surface layer. Large changes in relative humidity induce apparent changes to the surface layer. Humidity and temperature data was taken from a laboratory in an adjoining building on an Omega OMYL-RH25 temperature/humidity sensor at a rate of 0.5 measurements/hour. On the timescale of weeks, the overall trend in thickness of the surface layer is a roughly linear increase of ~0.05 nm per/week. The precise source of the thickness increase is unknown but we hypothesize the possibility of chemical change in the diamond surface or adventitious carbon deposition or a combination thereof. Regardless, the measurable change in surface properties with time (and humidity) motivate a more detailed understanding of diamond surface chemistry and stability.

To investigate this further the apparent change in thickness of $Al_2O_3$ is quantified as ΔDMAI and ΔH$_2$O where the thickness on either side of the plateau is subtracted from the plateau, Figure S3. In a typical ALD process, both of these values should be positive as the plateau before the oxidant pulse is the highest point. The change in apparent thickness caused by the DMAI pulse is roughly constant for steady-state growth but is totally inhibited during the nucleation of $Al_2O_3$ on the H-terminated diamond surfaces, Figure S4.

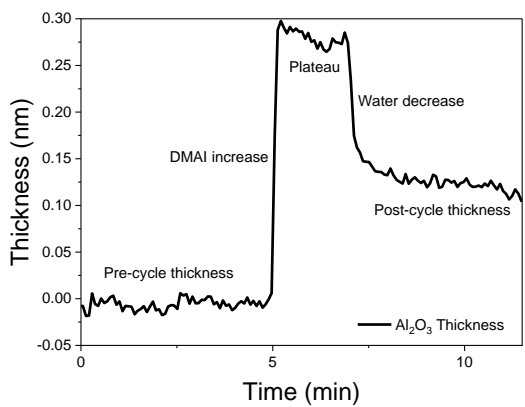

**Figure. S3.** Breakdown of the components of 1 ALD cycle in ellipsometry.

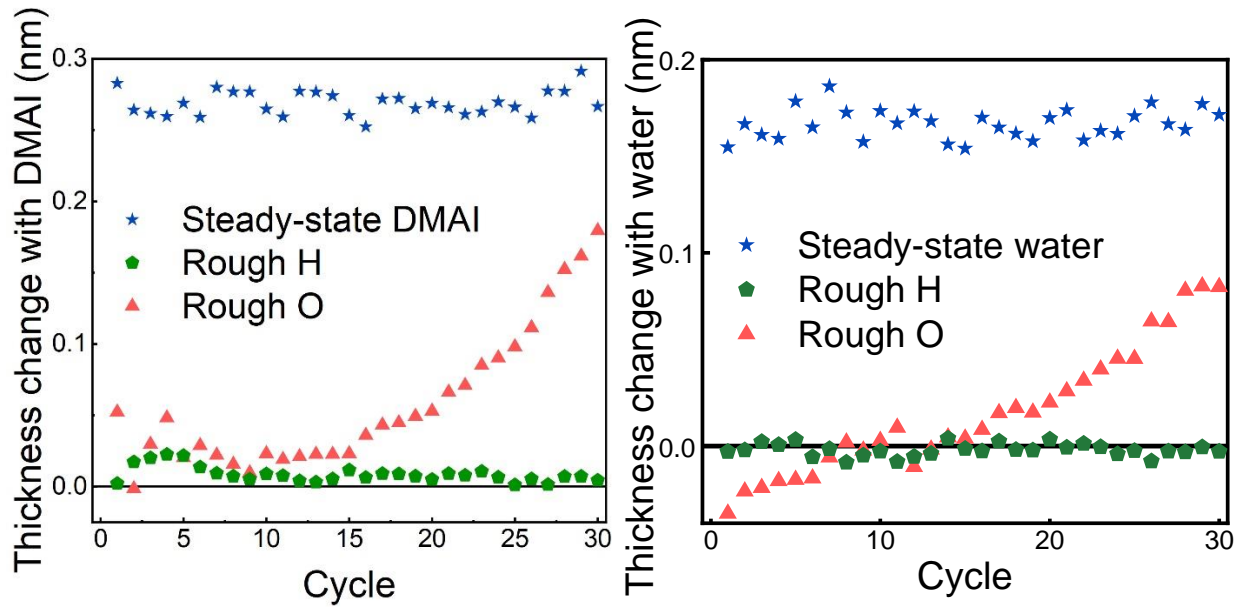

**Figure. S4** Change in apparent thickness due to DMAI (left) and water (right) pulses, line at 0 added as a guide to the eye.

When the ellipsometric data for the rough oxygen terminated surface is investigated in detail, progressively more DMAI adsorbs on the surface with increasing number of cycles. This could be consistent with island growth, as modeled by Nilsen et al, where the precursor adsorbs to the surface and the island grows vertically and laterally.[1] This however could also be consistent with the model proposed by Parson et al where progressively more nucleation sites are generated during the ALD process, possibly due to the surface's susceptibility to reaction with water.[2] During the first 13 cycles the change due to water on the rough oxygen terminated surface causes an increase in apparent thickness on the surface (a negative difference between the plateau and the following thickness), opposite of the expected effect! This would indicate addition of material to the surface, which is consistent with the observations in Figure S2 showing that triacid cleaned diamond is susceptible to modification by water vapor. This might also explain the shape of the ellipsometric data during these first 13 cycles (Figure 2).

The ratio between change in thickness caused by DMAI and water pulses should be the same as the film nucleates assuming that the same process is occurring. For example, a 2:1 ratio of optical thickness change is observed for the DMAI and water ALD process on itself. This 2:1 ratio is observed during the entire nucleation and early growth process on Si, however, on rough triacid treated diamond, the ratio did not consistently settle in to 2:1 until after cycle 15, suggesting that ALD according to the expected growth mechanism does not occur during early cycles. On samples that never nucleated, such as the rough H terminated surface the ratio never settles into a 2:1 ratio and has much larger values due to the changes in the thickness buried in the noise.

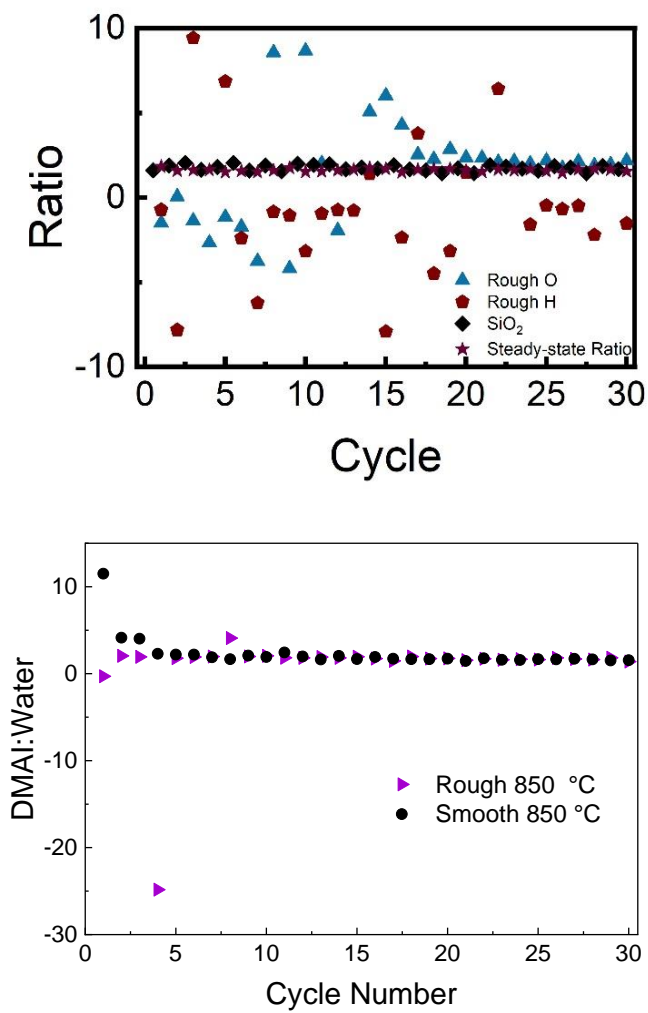

**Figure. S5.** Ratio of change in thickness due to DMAI and water for (top) $Al_2O_3$ on itself, nucleating on Si, triacid cleaned, and hydrogen plasma terminated rough diamond and (bottom) for samples annealed in high vacuum without atmospheric exposure.

The growth rate of the ALD process can be measured by the change in the measured thickess after each cycle. In the high vacuum annealed samples it is observed that the growth rate rapidly increases to the stead state of 0.1 nm/cycle in the first 10 cycles, consistent with the nucleation delay of 10 cycles observed. However, the growth rate of the atmospheric exposed samples was still increasing to the steady state rate after 30 cycles, demonstrating a marked difference in nucleation on the surface after exposure of the surface to atmosphere.

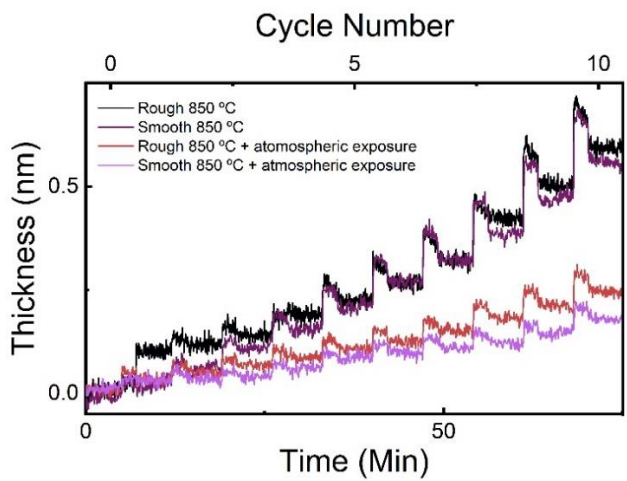

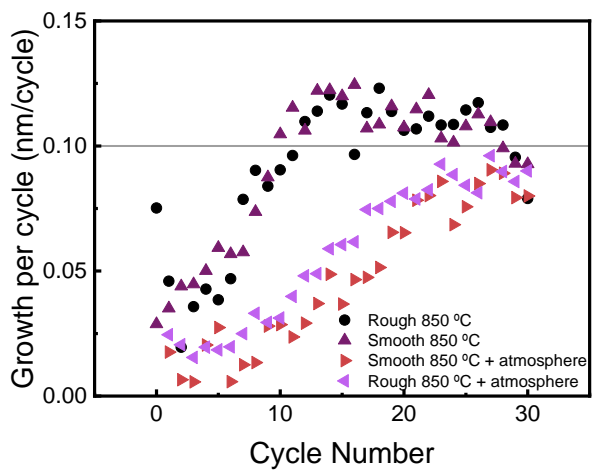

**Figure. S6** (Top) zoom of the first 10 cycles of the thickness of $Al_2O_3$ samples triacid cleaned and annealed to 850 °C before either ALD nucleation or atmospheric exposure before ALD nucleation. (Bottom) $Al_2O_3$ optical thickness increase per cycle. Line at 0.1 nm/cycle is a guide to the eye at the steady-state growth rate.

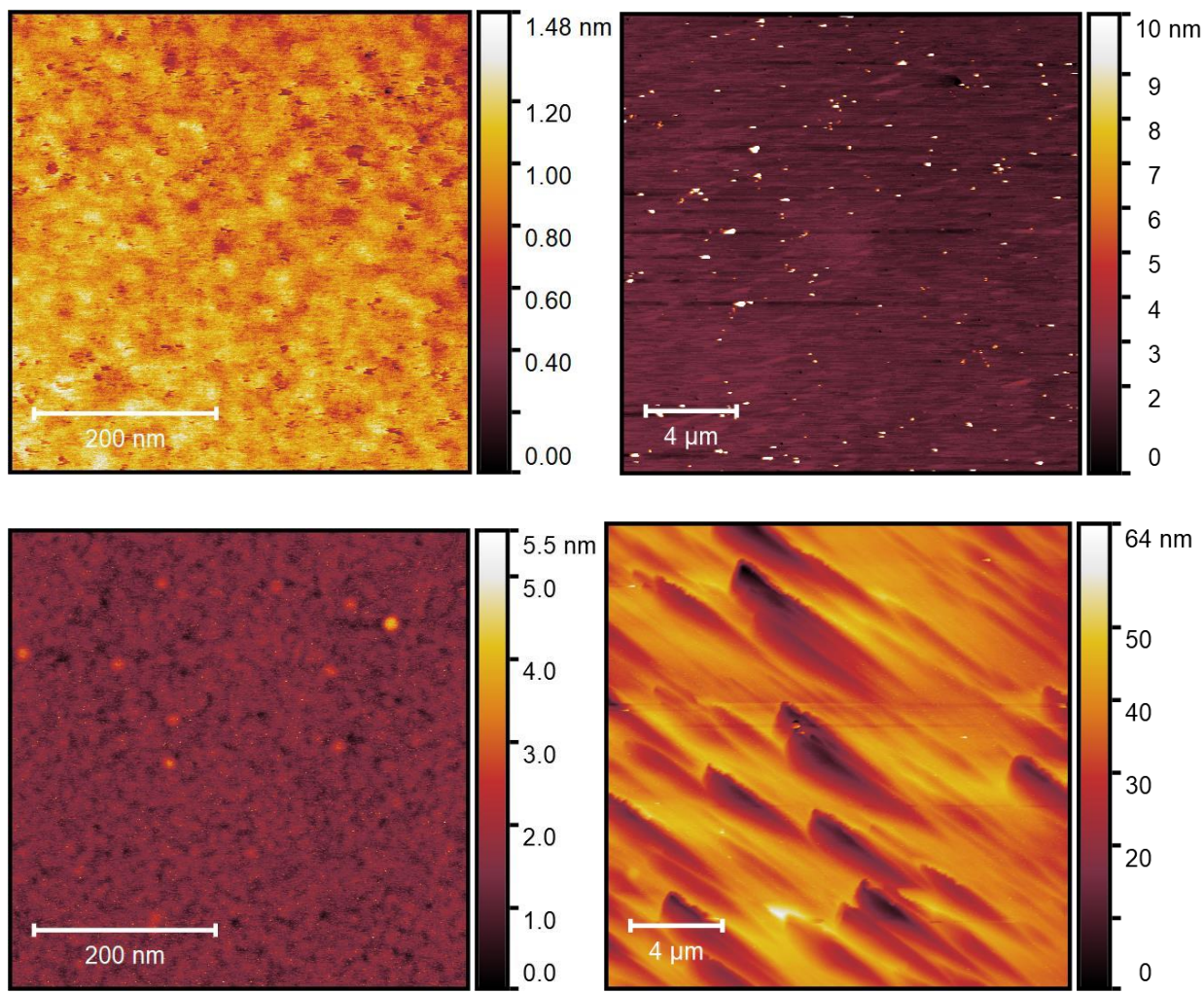

**Figure. S7** AFM images taken after 850 °C annealing in high vacuum and 30 cycles ALD deposition (top) smooth (bottom) rough surface. Showing minimal increase in roughness from annealing in high vacuum to 850 °C by AFM, and smooth conformal ALD films deposited on the surface.

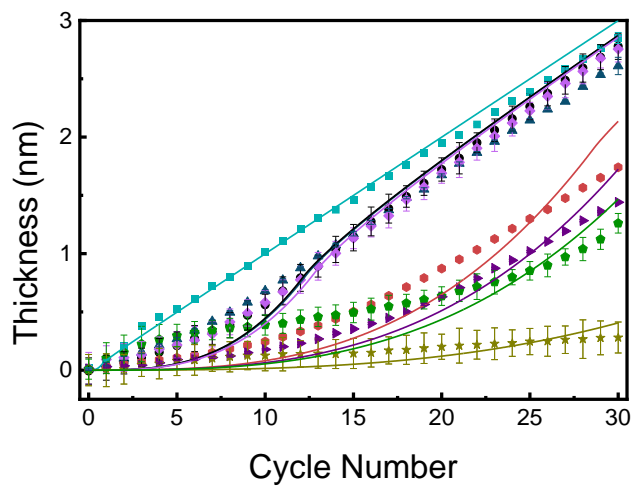

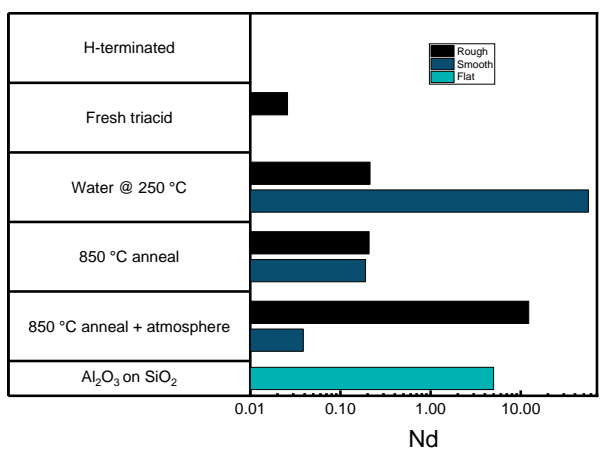

**Figure. S9** (Top) Thickness of $Al_2O_3$ with respect to cycle fit to the Nilsen model of nucleation density and (bottom) bar graph of the calculated nucleation density. Nd displayed on log scale to spawn range of nucleation densities. Nucleation density was not calculated for samples that did not ultimately nucleate in the 30 cycles of ALD performed.

REFFERENCES